\documentclass[final]{aipproc}

\layoutstyle{8x11double}

\usepackage{amsfonts,amssymb,amsmath,bm}
\usepackage{graphicx}

\newcommand{\be}{\begin{equation}}
\newcommand{\bea}{\begin{eqnarray}}
\newcommand{\ee}{\end{equation}}
\newcommand{\eea}{\end{eqnarray}}

\def\gb{\bm{\Gamma}}

\begin{document}

\title{A dynamical study of the Kugo-Ojima function}


\author{Daniele Binosi}{
  address={European Centre for Theoretical Studies in Nuclear Physics and Related Areas (ECT*), Villa Tambosi, Strada delle Tabarelle 286, I-38123 Villazzano (TN), Italy}
}

\begin{abstract}
 As has been recently realized, a certain two-point function $\Lambda_{\mu\nu}$ -- and its associated form factors $G$ and $L$ -- play a prominent role in the PT-BFM formulation of the Schwinger-Dyson equations used to study {\it gauge-invariantly} the gluon and ghost propagators. After showing that in the (background) Landau gauge $\Lambda_{\mu\nu}$ fully constrains the QCD ghost sector, we show that $G$ coincides with the Kugo-Ojima function $u$, whose infrared behavior has traditionally served as the standard criterion for the realization of the Kugo-Ojima confinement mechanism. The determination of the behavior of $G$ for all momenta through a combination of the available lattice data on the gluon and ghost propagators, as well as the dynamical equation $G$ satisfies, will be then discussed. In particular we will show that in the deep infrared the function deviates considerably from the value associated with the realization of the Kugo-Ojima confinement scenario; the dependence on the renormalization point of $u$, and especially of its value at $q^2=0$, will be also briefly discussed.
 \end{abstract}

\maketitle


Over the last few years large volumes ab-initio lattice gauge theory computations have established beyond any reasonable doubt that the gluon propagator and the ghost dressing function of pure Yang-Mills theories in the Landau gauge saturates in the deep infrared (IR) at a finite, non-vanishing value both for SU(2)~\cite{Cucchieri:2007md} and SU(3)~\cite{Bogolubsky:2007ud} gauge groups. 
Specifically choosing the $R_\xi$ Landau gauge and defining the gluon propagator cofactor $\Delta$, and the ghost dressing function $F$ as
\be
\Delta_{\mu\nu}(q)=-\mathrm{i}P_{\mu\nu}(q)\Delta(q^2), \qquad
D(q^2)=\mathrm{i}\frac{F(q^2)}{q^2},
\label{def}
\ee
where $P_{\mu\nu}(q)=g_{\mu\nu}-q_\mu q_\nu / q^2$ is the transverse projector, and $D(q^2)$ the ghost propagator, the aforementioned lattice results tell us that (in Euclidean space)
\be
\Delta^{-1}(0)>0,\qquad \mathrm{and}\qquad F(0)>0.
\ee

In the continuum formulation, the only way of obtaining these so-called {\it massive} solutions in a gauge invariant way and without breaking (either explicitly or softly) the BRST symmetry of the original Yang-Mills action, is within the PT-BFM framework~\cite{Aguilar:2008xm}, where 
a truncation scheme that respects gauge invariance at every level of the {\it dressed-loop} expansion has been developed in~\cite{Binosi:2007pi} using the systematic rearrangement of the entire Schwinger-Dyson series allowed by the pinch technique~\cite{Cornwall:1981zr,Cornwall:1989gv,Binosi:2002ft}.  

\begin{figure}[!t]
\includegraphics[scale=0.57]{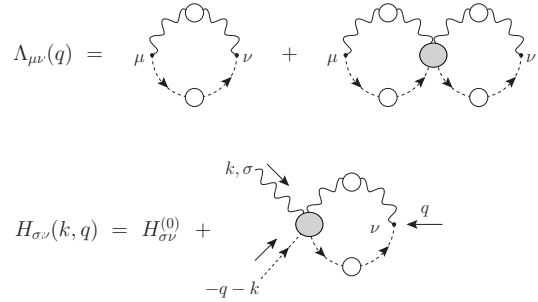}
\caption{\label{fig1}The auxiliary functions $\Lambda_{\mu\nu}$ and $H_{\sigma\nu}$ appearing in the PT-BFM framework.}
\end{figure}

In the PT-BFM construction one studies the PT-BFM propagator $\widehat{\Delta}$ which is related to the conventional propagator of Eq.~(\ref{def}) through the background-quantum identity~\cite{Grassi:1999tp}
\be
\widehat{\Delta}(q^2)=[1+G(q^2)]^2\Delta(q^2),
\label{bqi}
\ee
where $G$ is the $g_{\mu\nu}$ form factor appearing in the Lorentz decomposition of the auxiliary Green's function 
$\Lambda_{\mu\nu}$ defined as (see Fig.~\ref{fig1})
\bea
\Lambda_{\mu \nu}(q) &=&g^2C_A
\int_k 
D(k+q)\Delta_\mu^\sigma(k)\, H_{\sigma\nu}(k,q)
\nonumber \\
&=&g_{\mu\nu} G(q^2) + \frac{q_{\mu}q_{\nu}}{q^2} L(q^2).
\label{LDec}
\eea
The function $H_{\mu\nu}$ appearing above (Fig.~\ref{fig1} again)
is in fact a familiar object, since it appears in the all-order Slavnov-Taylor identity
satisfied by the standard  three-gluon vertex. It is also related to the full gluon-ghost vertex $\gb_{\mu}$ by the identity
$q^\nu H_{\mu\nu}(k,q)=-\mathrm{i}\gb_{\mu}(k,q)$; at tree-level, $H_{\mu\nu}^{(0)} =\mathrm{i}g_{\mu\nu}$ and $\gb^{(0)}_{\mu}(k,q)=\Gamma_\mu(k,q)=-q_\mu$. 

\begin{figure}[!t]
\includegraphics{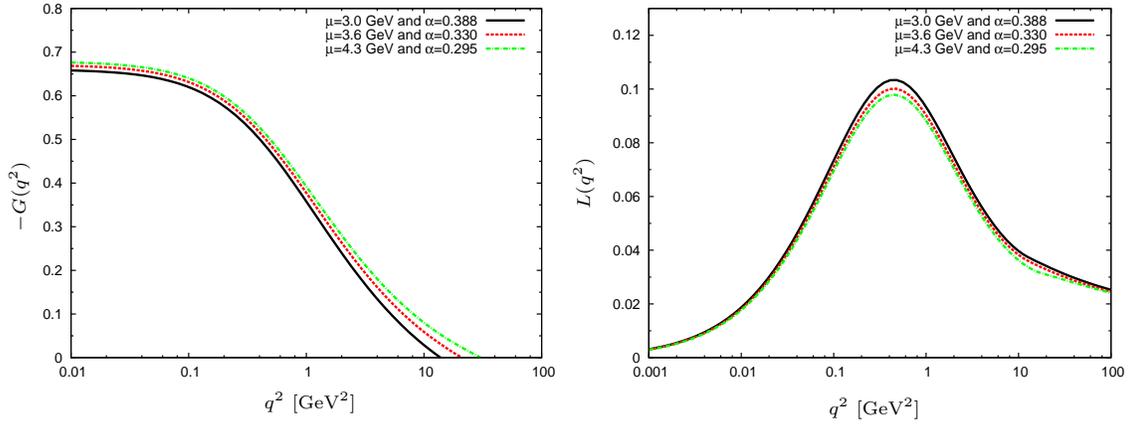}
\caption{\label{fig2}{\it Left panel}: The form factor $-G(q^2)$ determined from Eq. (24) at different renormalization points $\mu$ through the procedure described in the text. {\it Right panel}: Same as in the previous panel but this time for the $L(q^2)$ form factor.}
\end{figure}

The $G$ and $L$ form factors play a prominent role in the (background) Landau gauge, where the presence of an extra local functional equation (the so-called antighost equation) implies the identity~\cite{Grassi:2004yq}
\be
F^{-1}(q^2)=1+G(q^2)+L(q^2),
\label{BRST-1}
\ee
a relation that is valid also in the conventional Landau gauge~\cite{Kugo:1995km}.
Since, under very general conditions on the gluon and ghost propagators, $L(q^2)\to0$ when $q^2\to0$ one has the IR relation $F^{-1}(0)=1+G(0)$. Thus, we see that a divergent -- or {\it enhanced}~\cite{Alkofer:2000wg} --  dressing function requires the condition $G(0)=-1$. To the practitioners, this latter condition will look suspiciously similar to the Kugo-Ojima (KO) confinement criterion which demands (as a necessary condition for confinement through the so-called {\it quartet mechanism}) that a certain function $u(q^2)$ (the KO function) acquires the IR value\linebreak $u(0)=-1$~\cite{Kugo:1979gm}. Indeed,  it is possible to show that $G$ is nothing but the KO function~\cite{Grassi:2004yq,Aguilar:2009pp}
\be
u(q^2)\equiv G(q^2).
\label{equiv}
\ee
Therefore the form factor $G$ encodes practically all relevant information on the IR dynamics of the ghost sector, and, at least partially, the gluon sector as well [through the identity~(\ref{bqi})]. 

Approximating the three point functions $\gb_\mu$ and $H_{\mu\nu}$ with their tree-level value (the first approximation on the ghost-gluon vertex being supported by lattice studies), the dynamical equations satisfied by $G$ and $L$ read
\bea
G(q^2)\!\! &=&\!\!\!\! \frac{g^2 C_{\rm {A}}}{3}\int_k \left[2+ \frac{(k \cdot q)^2}{k^2 q^2}\right]\!\Delta (k)  D(k+q),
\nonumber\\
L(q^2)\!\! &=&\!\!\!\! \frac{g^2 C_{\rm {A}}}{3}\int_k \left[1- 3 \, \frac{(k \cdot q)^2}{k^2 q^2}\right]\!\Delta (k)  D(k+q).
\label{simple}
\eea 
Then, since within our approximation scheme, in the equations above only $\Delta$, $F$ (through the ghost propagator $D$) and $g$ appears, in order to determine the behavior  of $G$ and $L$ one can fully exploit the available lattice data on $\Delta$ and $F$ , through the following general strategy~\cite{Aguilar:2009pp,Aguilar:2010gm}. 
One starts by using the lattice gluon propagator as an input for the ghost SDE; then solves for the ghost dressing function, tuning the coupling constant $g$ such that the solution gives the best possible approximation to the lattice results. Obviously one must check that the coupling so obtained (at the renormalization scale used for the computation) is fully consistent with known perturbative results (obtained in the MOM scheme, which is the scheme used in our computations); this is indeed what happens~\cite{Aguilar:2009pp,Aguilar:2010gm}. 
At this point the one has the three building blocks $\Delta$,  $F$ and $g$ fully determined, and can start analyzing other quantities constructed from them such as the $G$ and $L$ form factors above~\cite{Aguilar:2009pp} or the renormalization group invariant effective charge~\cite{Aguilar:2010gm,Aguilar:2009nf}.

Before solving numerically the equations~(\ref{simple}), there is one last issue that needs to be addressed. Specifically, one needs to identify a renormalization procedure for $G$ and $L$ that does not break the identity (\ref{BRST-1}), which, due to its BRST origin, should not be deformed (within the PT-BFM scheme) by the renormalization process.
Note in fact that Eq.~(\ref{BRST-1}) constrains the cutoff-dependence 
of the unrenormalized quantities involved;
specifically, denoting by $Z_c$ the ghost wave-function renormalization constant ($Z_cF^{-1}_0=F^{-1}$) and with $Z_\Lambda$ the (yet unspecified) renormalization constant of the function $\Lambda_{\mu\nu}(q)$, with $Z_\Lambda[g^{\mu\nu}+\Lambda^{\mu\nu}_0]=g^{\mu\nu}+\Lambda^{\mu\nu}$, one finds that (\ref{BRST-1}) is preserved iff $Z_\Lambda=Z_c$~\cite{Aguilar:2009nf};  as a result, one finds
the relation
\bea 
& & Z_c(\Lambda^2, \mu^2)[1+G_0(q^2,\Lambda^2)+L_0(q^2,\Lambda^2)] \nonumber \\
& & \hspace{2cm} = 1+G(q^2,\mu^2)+ L(q^2,\mu^2).
\label{Zcren}
\eea
Imposing then the renormalization condition $F(\mu^2)=1$, 
going to Euclidean space, setting $q^2=x$, $k^2=y$ and $\alpha_s=g^2/4\pi$, 
and implementing the standard angular approximation,  
one finds the renormalized equations~\cite{Aguilar:2009pp,Aguilar:2009nf}
\bea
1+G(x) &=&  Z_c - \frac{\alpha_s C_{\rm {A}}}{16\pi}\left[
\frac{F(x)}{x}\int_{0}^{x}\!\!\! dy\  y \left(3 + \frac{y}{3x}\right) \Delta(y) \right.
\nonumber \\ 
&+&\left. \int_{x}^{\infty}\!\!\! dy \left(3 + \frac{x}{3y}\right)\Delta(y)F(y) 
\right],
\nonumber\\
L(x) &=&  \frac{\alpha_s C_{\rm {A}}}{12\pi} \left[
\frac{F(x)}{x^2}\int_{0}^{x}\!\!\! dy\ y^2 \Delta(y) \right.
\nonumber \\ 
&+&\left. x \int_{x}^{\infty}\!\!\! dy \frac{\Delta(y) F(y)}{y}
\right].
\label{FGL}
\eea
Notice that $L$ is finite, as expected from power counting; in addition, we see (by means of the change of variables $y = zx$) that if $\Delta$ and $F$ are IR finite, then $L(0) = 0$, as mentioned before (notice however that the same result is obtained for {\it scaling solutions}~\cite{Alkofer:2000wg}, where $\Delta(y) \sim y^{a}$ and  $F(x)\sim x^{b}$, provided that $a+b>-1$).

At this point, all necessary ingredients for determining the functions $G$ and $L$ are available. Substituting them into the corresponding equations given in (\ref{FGL}), we obtain the solutions shown in Fig.~\ref{fig2}, where we see that $L$ is subdominant, and that indeed it vanishes in the deep IR.
Also, the $\mu$ dependence of the KO function and the KO parameter are clearly shown; for the range of renormalization points $\mu$ chosen, the KO function saturates in the deep IR to the value $G(0)\approx-0.6$ which deviates irremediably from the value $G(0)=-1$ required for the realization of the KO confinement scenario.

\begin{figure}[!t]
\includegraphics{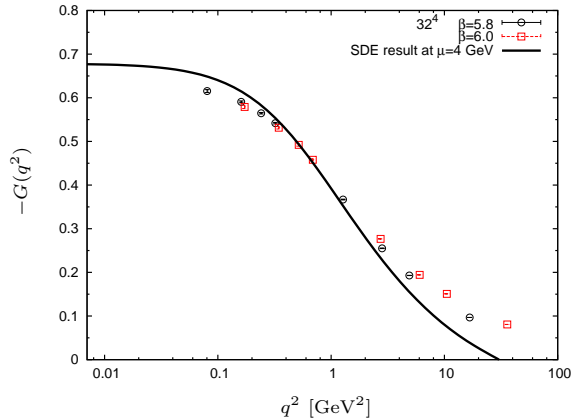}
\caption{\label{fig3}Comparison between our results and direct lattice calculation of the KO function $-u(q^2)$ at $\mu=4$ GeV.}
\end{figure}

The curves plotted for $G$ should finally be compared with those obtained on the lattice in~\cite{Sternbeck:2006rd}, where the KO function $u$ was studied in terms of Monte Carlo averages, and its asymptotic behavior inferred from the identity (\ref{BRST-1}). Even though in~\cite{Sternbeck:2006rd} the extrapolation towards the zero limit was problematic, due to a lack of knowledge of the function $L$ (our analysis does not suffer from such a limitation, given that $L$ is completely determined by its own equation) and that, for essentially the same reason, the renormalization procedures employed are different, we clearly see in Fig.~\ref{fig3} the same behavior emerging, and in particular the saturation of the KO function in the deep IR to a value sensibly different from the critical $-1$.

In conclusion, we have shown that the massive gluon propagator $\Delta$ and ghost dressing function $F$ found as solutions to the SDE and confirmed by all large volume lattice simulations up to now, do not support a confinement scenario based on the original KO mechanism/criterion. However, due to the equality (\ref{equiv}) between the KO function $u$ and the auxiliary function $G$, and the central role that the latter function has in the PT-BFM scheme, it would still be very interesting to carry a thorough study of such function on large volume lattices (for different space-time dimensions and gauge groups). \\

\noindent{\it Acknowledgments.} The author thanks the organizers of {\it Quark Confinement IX International Conference} for the hospitality and the very stimulating conference.

\bibliographystyle{aipproc}

\begin{thebibliography}{99}

\bibitem{Cucchieri:2007md}
  A.~Cucchieri and T.~Mendes,
  PoS {\bf LAT2007}, 297 (2007);
  Phys.\ Rev.\ Lett.\  {\bf 100}, 241601 (2008);
  Phys.\ Rev.\  D {\bf 81}, 016005 (2010).

\bibitem{Bogolubsky:2007ud}
   I.~L.~Bogolubsky, E.~M.~Ilgenfritz, M.~Muller-Preussker and A.~Sternbeck,
  PoS {LATTICE}, 290 (2007);
  Phys.\ Lett.\  B {\bf 676}, 69 (2009);
  O.~Oliveira and P.~J.~Silva,
  PoS {\bf LAT2009}, 226 (2009).

\bibitem{Aguilar:2008xm}
  A.~C.~Aguilar, D.~Binosi and J.~Papavassiliou,
  Phys.\ Rev.\  D {\bf 78}, 025010 (2008);
  Phys.\ Rev.\  {\bf D81}, 125025 (2010).

\bibitem{Binosi:2007pi}
  D.~Binosi and J.~Papavassiliou,
  Phys.\ Rev.\  D {\bf 77}, 061702(R) (2008);
  JHEP {\bf 0811}, 063 (2008).
  


\bibitem{Cornwall:1981zr}
  J.~M.~Cornwall,
  Phys.\ Rev.\  D {\bf 26}, 1453 (1982).


\bibitem{Cornwall:1989gv}
  J.~M.~Cornwall and J.~Papavassiliou,
  Phys.\ Rev.\  D {\bf 40}, 3474 (1989).

\bibitem{Binosi:2002ft}
  D.~Binosi and J.~Papavassiliou,
  Phys.\ Rev.\  D {\bf 66}(R), 111901 (2002);
  J.\ Phys.\ G {\bf 30}, 203 (2004);
  Phys.\ Rept.\  {\bf 479}, 1 (2009).

\bibitem{Grassi:1999tp}
  P.~A.~Grassi, T.~Hurth and M.~Steinhauser,
  Annals Phys.\  {\bf 288}, 197 (2001);
  D.~Binosi and J.~Papavassiliou,
  Phys.\ Rev.\  D {\bf 66}, 025024 (2002).

\bibitem{Grassi:2004yq}
  P.~A.~Grassi, T.~Hurth and A.~Quadri,
  Phys.\ Rev.\  D {\bf 70}, 105014 (2004).

\bibitem{Kugo:1995km}
  T.~Kugo,
  arXiv:hep-th/9511033.

\bibitem{Alkofer:2000wg}
  R.~Alkofer, L.~von Smekal,
  Phys.\ Rept.\  {\bf 353}, 281 (2001); 
  C.~S.~Fischer,
  J.\ Phys.\ G {\bf 32}, R253 (2006).


\bibitem{Aguilar:2009pp}
  A.~C.~Aguilar, D.~Binosi and J.~Papavassiliou,
  JHEP {\bf 0911}, 066 (2009).

\bibitem{Kugo:1979gm}
  T.~Kugo and I.~Ojima,
  Prog.\ Theor.\ Phys.\ Suppl.\  {\bf 66}, 1 (1979).
  
\bibitem{Aguilar:2010gm}
  A.~C.~Aguilar, D.~Binosi, J.~Papavassiliou,
  JHEP {\bf 1007}, 002 (2010).

\bibitem{Aguilar:2009nf}
  A.~C.~Aguilar, D.~Binosi, J.~Papavassiliou and J. Rodriguez-Quintero,
  Phys.\ Rev.\  {\bf D80}, 085018 (2009).

\bibitem{Sternbeck:2006rd}
  A.~Sternbeck,
  arXiv:hep-lat/0609016.



\end{thebibliography}

\end{document}